\documentclass[%
letterpaper,
]{article}

\makeatletter
\def\switch@array{}
\makeatother

\usepackage{graphicx}
\usepackage{dcolumn}
\usepackage{bm}
\usepackage{verbatim}

\usepackage[english]{babel}
\usepackage{achemso}

\usepackage{physics}
\usepackage{xcolor}
\usepackage{textcomp}
\usepackage{authblk}

\graphicspath{{./images/}{}}

\newif\iflinksupp
\linksuppfalse         

\iflinksupp
  \usepackage{xr-hyper}
  \newcommand{\suppref}[2]{supplementary \Cref*{sup-#2}}
  \newcommand{\supprefs}[3]{supplementary \Cref*{sup-#2,sup-#3}}
\else
  \newcommand{\suppref}[2]{supplementary #1} 
  \newcommand{\supprefs}[3]{supplementary #1} 
\fi

\usepackage{hyperref}
\hypersetup{pdfstartview={FitH},pdfpagemode={UseNone}, breaklinks=true,colorlinks=true,allcolors=blue,bookmarksopen=true, pdfnewwindow=true}
\usepackage[all]{hypcap}

\usepackage{cleveref}
\Crefname{figure}{Fig.}{Figs.}
\Crefname{equation}{Eq.}{Eqs.}
\crefname{supplement}{supplementary}{supplementaries}
\Crefname{supplement}{Supplementary}{Supplementaries}

\newcommand{\vc}[1]{\va{#1}}
\newcommand{\iprod}[1]{\langle#1,#1\rangle}
\newcommand{\iprodM}[2]{\langle#2,#1#2\rangle}
\newcommand{\vuw}[1]{\vc{\mathit{#1}}}

\begin{document}


\title{
Chiral lasing via broken parity-time symmetry
in bound-state-in-the-continuum metasurfaces}

\author[1]{Matthew Parry}
\date{*Email: Matthew.Parry@anu.edu.au}

\author[1]{Daria A. Smirnova}
\author[1]{Andrey A. Sukhorukov}
\author[1]{Dragomir N. Neshev}
\affil[1]{ARC Centre of Excellence for Transformative Meta-Optical Systems (TMOS), Department of Electronic Materials Engineering, Research School of Physics, Australian National University, Canberra, ACT 2600, Australia}


\maketitle

\begin{abstract}
We propose a concept for chiral lasing from planar metasurfaces that obviates the need for traditional out-of-plane symmetry breaking by exploiting spatial gain-loss modulation to break parity-time symmetry. We explain the underlying non-Hermitian physics of this design principle using a coupled-mode model of a four-site plaquette. The symmetry requirements for such chiral emission are explained with a general symmetry analysis based on projection operator matrices, implemented algorithmically for automated evaluation. This method enables the design of planar metasurfaces capable of emitting nearly-pure circularly polarized light. We apply our analysis to simulations of both symmetric and asymmetric versions of a Fylfot metasurface design and demonstrate that the gain mode at the parity-time symmetric exceptional point exhibits chiral emission. Lastly, we present a readily manufacturable metasurface made from an InGaAs slab, showing that such a metasurface laser can be actively tuned from linear to circular polarization.

\end{abstract}


\section{Introduction}
The geometric symmetry of a photonic system encompasses the symmetry of both the permittivity distribution and the crystal structure, or nonlinear tensor, as well as the symmetry of the electromagnetic fields and eigenmodes~\cite{Reinke:2011-66603:PRE}. Consideration of these symmetries in two-(2D) or three-dimensions (3D) is essential when designing a photonic system and geometric symmetry control can be used to create effects such as flat dispersion through the engineering of Dirac cones,~\cite{Huang2011,Sakoda:2012-25181:OE,Zhen2015}, bound-states-in-the-continuum (BICs)~\cite{Kilic2008,Hsu:2016-16048:NRM} and ultra-high-Q Metasurface (MS) lasers~\cite{Song:2021-31105:APL} among many other applications. 
In particular, the 3D symmetries are ultimately important to define chirality in applications such as chiral sensing~\cite{Barkaoui:LPR:2025} and chiral lasing~\cite{Zhang:Sci:2022}.

\begin{figure*}[ht!]
    \centering
    \includegraphics[width=0.7\textwidth]{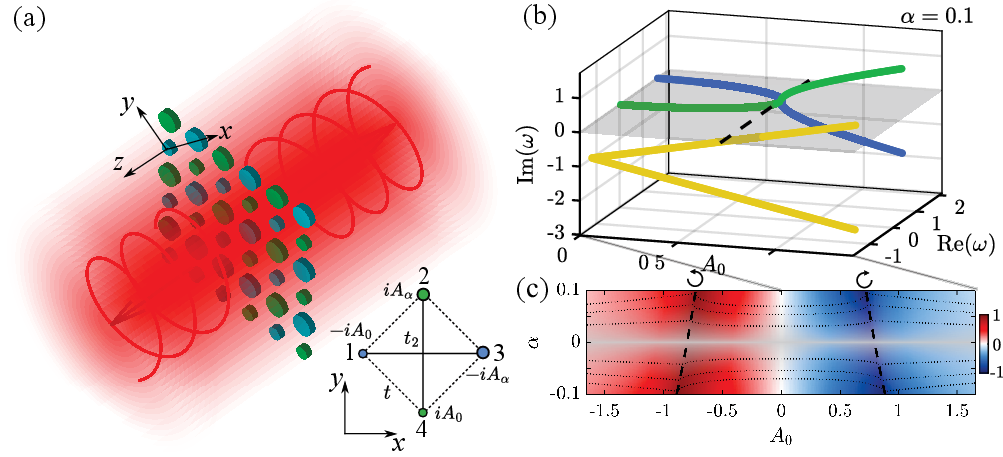}
    \caption{
    (a) Schematic of a metasurface emitting circularly polarized light. Low inset: lattice model for a quadrumer with nearest-neighbor and next-nearest-neighbor couplings, $t$ and $t_2$, respectively, and on-site gain–loss modulation. Sites 2 and 3 are perturbed, quantified by the asymmetry parameter $\alpha$.
    (b) Complex-valued spectra of the four modes of a quadrumer, with one mode brought above the $\mathrm{Im}(\omega)=0$ plane into the gain region in the vicinity of the exceptional point (EP) at $\alpha=0.1$, $t=0.4$, $t_0=2$, $t_2 = \gamma = 1$.
    (c) Corresponding map of the Stokes parameter $S_3$, characterizing the radiative admixture for this mode in the plane of asymmetry parameters $(A_0,\alpha)$. The dashed black line indicates the position of the EP. The color scale represents the degree of circular polarization $S_3$, while the transparency encodes the total radiated intensity via a power-law scaling $S_0^{1/4}$; contours of constant $S_0$ are overlaid with dotted lines.
    }
    \label{fig:fig_Concept}
\end{figure*}

The common wisdom is that achieving chiral emission at the normal out-of-plane direction from a thin metasurface requires breaking the out-of-plane symmetry~\cite{Hu:AdvPhotRes:2025}, for instance by introducing a substrate or employing oblique patterning~\cite{Chen:Nature:2023}. Here, we propose a fundamentally different approach to realizing chiral emission from a planar metasurface, enabled by symmetry engineering of the in-plane distribution of optical gain and loss. We show that this mechanism is fundamentally rooted in non-Hermitian physics and the interplay of coupled collective modes, and thus it is naturally generalizable to other platforms, including polaritonic lattices and quantum systems governed by complex potentials.

We consider Rotation-Time (\(\mathcal{RT}\)) symmetry~\cite{Ge:2014-31011:PRX,Ge:2015-62135:PRA} specifically, which can be referred to as Parity-Time (\(\mathcal{PT}\)) symmetry~\cite{Wang:2023-442:ADOP,Ozdemir:2019-783:NMAT,Feng:2017-752:NPHOT} if we use a more generalized definition of \(\mathcal{PT}\) symmetry~\cite{Wang:2023-1443:JOSB} as
\begin{equation}
    [\mathcal{OT},\mathcal{H}] = 0,
\end{equation}
for some unitary operator \(\mathcal{O}\) such that \((\mathcal{OT})^2=1\) and where \(\mathcal{H}\) is the Hamiltonian. In this letter, we will substitute a \(\flatfrac{\pi}{2}\) rotation symmetry (\(\mathcal{R}_{\flatfrac{\pi}{2}}\)) for the parity operation to achieve the same effects as a standard parity operator.~\cite{Ruter:2010-192:NPHYS,Guo:2009-93902:PRL} 

In the resulting broken \(\mathcal{PT}\) symmetry phase, one of the pair of BIC modes is a gain mode which could serve as the lasing mode. For this we require a high-quality factor mode to aid the lasing and so we will take advantage of quasi-bound States in the Continuum (quasi-BICS) by introducing an asymmetry into the MS.~\cite{Kilic2008,Koshelev:2018-193903:PRL} 


The concept is illustrated in \Cref{fig:fig_Concept}(a) which shows an abstract metasurface composed of a square lattice of quadrumers, each consisting of four elements with alternating gain and loss of varying magnitude, $A_0$. We focus here on quasi-TE polarization --- i.e., on modes characterized by an in-plane scalar field distribution of the out-of-plane magnetic field $H_z$ and an in-plane electric field vector. 
At vanishing $A_0=0$, each quadrumer supports four modes that can be classified by symmetry: two dipolar modes ($E_{\pm}$), which can couple to plane waves by symmetry, and two non-radiative symmetry-protected BIC modes ($\ A_1$ and $B_2$) (See \suppref{Section~S1}{sec:eff-hamiltonian}). Upon increasing $A_0$, a pair of non-radiative modes gives rise to the exceptional point in the complex-valued spectrum.
We developed a lattice model of the mode couplings in our $\mathcal{R}_{\flatfrac{\pi}{2}}\mathcal{T}$ system, as illustrated in the inset of \Cref{fig:fig_Concept}(a), both in the real-space basis—where the site amplitudes can be treated as $H_z$ field components—and under its transformation to the mode basis, $\begin{bmatrix} E_+, & E_-, & A_1, & B_2 \end{bmatrix}$.
 
Thus, our approach is to break $\mathcal{RT}$ symmetry so that it predominantly admixes one of the circularly polarized dipolar modes ($E_{\pm}$) to the high-quality mode that acquires gain. For this, we slightly perturb the sites 2 and 3.
This will facilitate chiral lasing, enabling emission of circularly polarized light in the normal (out-of-plane) direction. 

It has previously been shown that \(\mathcal{PT}\) symmetry can be used to create MS lasers,~\cite{Brandstetter:2014-4034:NCOM,Peng:2014-328:SCI,Gu:2016-588:LPR,Hodaei:2014-975:SCI} but \(\mathcal{PT}\) symmetry by itself is not sufficient to achieve chiral lasing. In addition, an important advance would be to be able to design MS lasers with arbitrary and actively tunable polarization.  This design process requires that symmetry analyses be performed on every aspect of geometric symmetry, not just \(\mathcal{PT}\) symmetry.  We need to analyze the symmetry of the permittivity distribution as well as that of the eigenmodes of the MS.  Because of the importance of methods such as inverse design, optimization routines, and machine learning, it is also essential that the symmetry analyses be able to be expressed in algorithmic form so that they can be incorporated into these design algorithms.  

We have previously shown that one can use projection operator matrices, or projectors, to measure the symmetry of scalar and vector fields.~\cite{Parry:2024-75406:PRB}  To do this, we represent the field as a vector, \(\vuw{u}\), and the projection operator is then a matrix that we apply to the vector to give the component of the vector that has the given symmetry.  We denote the projectors as
\begin{equation}
    \hat{B}_1^{C2v},
\end{equation}
where the main letter and subscript denote the irreducible representation and the superscript the point group -- In this case, \(C_{2v}\ B_1\).  We can then numerically quantify the degree to which a field has a given symmetry with a symmetry parameter defined as a ratio of inner products as
\begin{equation}\label{eq:sym-param}
    \eta = \frac{\iprodM{\hat{B}_1^{C2v}}{\vuw{u}}}{\iprod{\vuw{u}}},
\end{equation}
where \(0\leq\eta\leq1\).

Projectors can also be used as a representation of the algebra of symmetry 
(See \supprefs{Sections~S3 and S4}{sec:proj-algebra}{sec:char-space}).
That is, for every relationship between the irreducible representations of different point groups, there is a corresponding equation for the projectors.  This means that the algebra of symmetry can be rendered in algorithmic form and incorporated into optimization~\cite{Parry:2024-75406:PRB} or machine learning routines with ease.

We begin with a symmetry analysis of the problem of creating chiral emission from a planar MS. For a MS without phase modulations that create orbital angular momentum, we can use end-fire coupling to predict the far-field polarization from the eigenmode field.~\cite{Overvig:2020-35434:PRB}  In such an approximation, the symmetry of the eigenmode field implies the same symmetry in the polarization of the emitted field.  By \emph{symmetry of the polarization} we mean that H polarization has the same symmetry as the linear function \(x\) and V has the symmetry of the linear function \(y\), with examples of scalar fields with these symmetries shown in 
\suppref{Fig.~S2}{fig:plot_sym}.  
For the case of circular polarization, which we seek, we have RCP (\(\mathrm{H}-i\mathrm{V}\)) having the symmetry of \(x-iy\) and LCP (\(\mathrm{H}+i\mathrm{V}\)) having the symmetry of \(x+iy\).  
These linear symmetries also correspond to particular irreducible representations, as outlined in \suppref{Table~S3}{tab:pol-sym} and \suppref{Section~S5}{sec:lin-sym}.  This then means that if an eigenmode has \(C_4\ E_{11}\) symmetry then, by the end-fire coupling approximation, it will emit LCP radiation.

Each of the uncoupled modes of an MS has the symmetry of one of the irreducible representations of the point group of the MS.  In our case we want LCP or RCP polarization which means that we want to add either an \(\hat{E}_{11}^{C4}\) or \(\hat{E}_{22}^{C4}\) symmetry component to the BIC, and hence we begin with a \(C_4\) symmetric MS (In 3D it is \(C_{4h}\), but a 2D \(C_4\) model is suitable for our purposes).  By saying that the MS has \(C_4\) symmetry, we mean that the permittivity distribution has \(\hat{A}^{C4}\) symmetry since \(A\) is the completely symmetric irreducible representation of \(C_4\).

We require \(\hat{E}^{C4}\) symmetry to be added to the BICs,~\footnote{The \(E\) representation of \(C_4\) is not really irreducible, but is normally treated as though it were.} but there is no guarantee that the \(\hat{E}^{C4}\) component will emit complex polarized light. This occurs because \(\hat{E}^{C4}\) encompasses all polarizations and hence we can not guarantee what the polarization will be. We need to create modes with \(\hat{E}_{11}^{C4}\) and \(\hat{E}_{22}^{C4}\) symmetry specifically (Note that \(\hat{E}^{C4} = \hat{E}^{C4v}\) and so the structure's symmetry does not have to be \(C_4\)).

To get around this problem, we note that the characters of the \(\hat{E}_{11}^{C4}\) and \(\hat{E}_{22}^{C4}\) symmetries are imaginary with respect to a \(\flatfrac{\pi}{2}\) rotation.  For the MS to have \(\hat{E}_{11}^{C4}\) symmetry would entail that the refractive index of each of the four sites of the lattice, moving in an anti-clockwise direction, be \(n\), \(in\), \(-n\) and \(-in\) respectively.  This, however, is not possible and so we try an alternating matched gain and loss in the lattice sites as shown in the inset of \Cref{fig:fig_Concept}a.

The underlying physics is captured by a coupled-mode theory (CMT) model (See \suppref{Section~S1}{sec:eff-hamiltonian}) with its $4 \times 4$ Hamiltonian in the real-space sublattice basis being 
\begin{equation}
\hat{H}=
\left[
\begin{smallmatrix}
- i A_0 - \frac{i\gamma}{2} & t & t_2 + \frac{i \gamma}{2} & t \\
t & i A_\alpha - 2 \alpha t_0 - \frac{i\gamma}{2} & t & t_2 + \frac{i \gamma}{2} \\
t_2 + \frac{i \gamma}{2} & t & - i A_\alpha - 2 \alpha t_0 - \frac{i \gamma}{2} & t \\
t & t_2 + \frac{i \gamma}{2} & t & i A_0 - \frac{i \gamma}{2}
\end{smallmatrix}
\right]. \label{eq:H_CMT}
\end{equation}
Here, $\gamma$ accounts for the radiative losses, $t_0$ quantifies the effect of the perturbation on the on-site frequency, and $\alpha \ll 1$ is a small symmetry-breaking parameter. 
The corresponding representative spectrum is shown in Fig.~\ref{fig:fig_Concept}(b). Fig.~\ref{fig:fig_Concept}(c) illustrates the polarization state of 
radiation for the gain mode, evaluated from the amplitudes of its dipolar components in the parameter plane $(A_0,\alpha)$, with the $S_3$ Stokes parameter approaching unity near the exceptional point (EP). Changing the sign of $A_0$ (swapping gain and loss) effectively flips the helicity.

In the mirror-symmetric gain-loss configuration, the net energy flow (the net angular momentum density) around the four-site plaquette 
averages to zero.
A tiny asymmetry favors helicity of flow, leading to the emergence of a 
near-field chiral vortex. 
This chiral near-field vortex Fourier-transforms into helical far-field radiation.

A simple yet insightful solution can be obtained from the eigenvalue problem 
$\hat{H} \bm{\psi} = \omega \bm{\psi}$ with parameters $t_2 = t_0 = \gamma = 0$. 
For convenience, we define
$
A_{0,\alpha} = A(1 \mp \alpha).
$
At the exceptional point $A_{\text{EP}} = {2t}/({1-\alpha^2})$, the state vector reads
\begin{equation}
\bm{\psi}_{\rm EP} \propto 
\left( 1,\, i,\, 1,\, i \right)^T 
+ \alpha \left( 1,\, -i,\, -1,\, i \right)^T.
\end{equation}
This state represents a superposition of $\pi/2$ phase-shifted dark modes (the first term) and a small, $\propto \alpha$, admixture of a circularly polarized dipole mode (the second term), which clearly exhibits the vortex pattern, 
giving rise to net helicity. 
For a four-site plaquette with uniform coupling $t$, the net circulation can be expressed as a loop sum:
\begin{equation}
\mathcal{C} 
= 2 t \sum_{n=1}^{4} | \psi_n | \, | \psi_{n+1} | \, \sin(\phi_{n+1} - \phi_n),
\end{equation}
where $\psi_n = |\psi_n| e^{i \phi_n}$ is the complex amplitude at site $n$, 
and site $5 \equiv 1$ closes the loop.
For small $\alpha$, the circulation of the exceptional-point vector is $\mathcal{C} \approx -8 t \alpha^2$, with the sign indicating the direction of circulation.
This solution is compact and simple compared to PT microrings with sophisticated shapes.~\cite{Zhang2020,Su2023,Ren2018,Miao2016}

The model~\eqref{eq:H_CMT} also suggests that gain–loss modulation lifts the degeneracy between the linearly-polarized dipoles (shown with yellow lines in Fig.~\ref{fig:fig_Concept}(b)) so that one can lase without symmetry breaking. Normally, this would require overcoming a threshold to compensate for radiative losses, but an \(\mathcal{R}_{\flatfrac{\pi}{2}}\mathcal{T}\) symmetry EP for the high-$Q$ states can occur at a much lower threshold. The suggested perturbation scheme realizes such a situation and in this latter scheme, with an EP, the threshold can be lower
(\Cref{fig:fig_Concept}b). This can assist with low-threshold circularly polarized lasing.

\begin{figure}[!tb]
    \centering
    \includegraphics[width=\columnwidth]{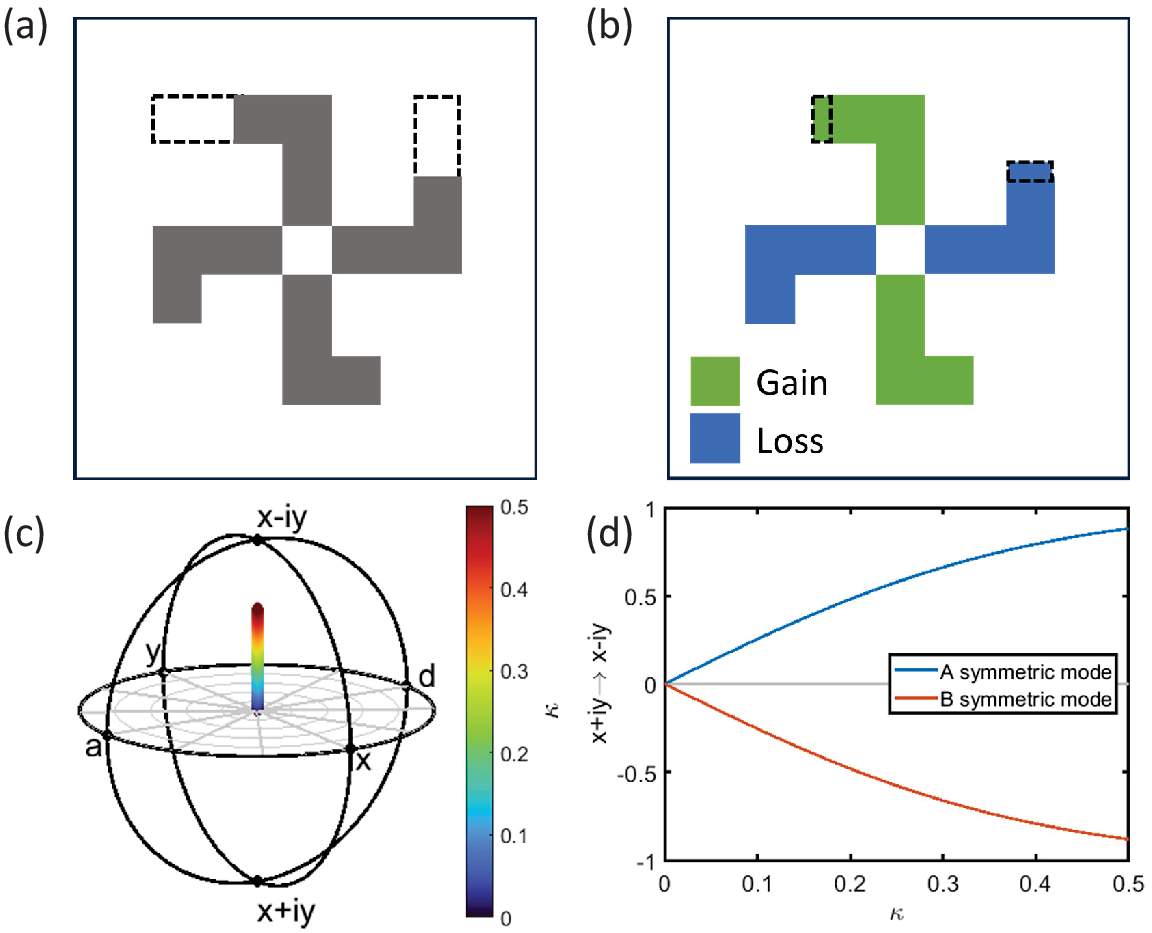}
	\caption[Fylfot Metasurface]{
    (a) \(C_4\ A\) symmetric fylfot metasurface design. 
    (b) The design for the metasurface showing alternating matched loss and gain. We have introduced asymmetry by extending two of the arms by 50~nm.
    (c) Linear symmetry of the asymmetric part of the permittivity distribution with the arms extended by 50~nm and increasing value of \(\kappa\), the imaginary part of the refractive index.  
    (d) Complex linear symmetry axis analysis for both \(\hat{A}^{C4}\) and \(\hat{B}^{C4}\) modes.}
    \label{fig:pt1}
\end{figure}

\section{Results and Discussions}
\subsection{Symmetric Fylfot Metasurface}
One possibility for realizing this model is the Fylfot unit cell shown in \Cref{fig:pt1}a.  We can add asymmetry to this MS~\cite{Kilic2008,Koshelev:2018-193903:PRL} by simply extending two of the arms as indicated by the dashed lines. We can test the symmetry of this MS using projection operators by creating a 2D representation of the permittivity distribution (with \(n\pm i\kappa\)) as shown in \Cref{fig:pt1}b. We represent this as a vector and by then using \Cref{eq:sym-param} with each of the projectors in \suppref{Table~S3}{tab:pol-sym} we can obtain a complete, six-dimensional quantification of the linear symmetry of the MS.  By projecting out half of these dimensions, we can plot the linear symmetry 
(See \suppref{Section~S7}{sec:plot})
for different values of \(\kappa\) as shown in \Cref{fig:pt1}c (The linear symmetry of the asymmetric part of the permittivity distribution only, as per the method of Ref.~\cite{Overvig:2020-35434:PRB}).  The six cardinal points represent the six types of linear symmetry, but it is important not to confuse this with a Poincar\'e sphere.  The plot only shows if there is a \emph{bias} towards \(x\) symmetry over \(y\), \(d\) over \(a\), \(x+iy\) over \(x-iy\) or visa versa, and does not show if linear symmetry exists or not.  That is, one could get a result of \((0,0,0)\) even though there is linear symmetry if there is no bias towards one type of linear symmetry over another.

From \Cref{fig:pt1}c we can see that as \(\kappa\) increases we get an increasing bias towards \(x-iy\) (i.e. \(\hat{E}_{22}^{C4}\)) symmetry as desired.  We can then follow the method of Ref.~\cite{Overvig:2020-35434:PRB} to find what the polarization of modes of different symmetry will be.  With the \(C_4\) point group there are two dark symmetries: \(A\), which is completely symmetric with respect to a \(\flatfrac{\pi}{2}\) rotation, such as an \(m_z\) dipole; and \(B\), which is anti-symmetric, such as an \(m_z\) quadrupole.  The results for these two symmetries are shown in \Cref{fig:pt1}d, which shows only the \(x+iy\rightarrow x-iy\) axis since the modes have no bias in the real linear symmetry plane.  The plot indicates that the two types of modes are biased towards opposite complex symmetries.

Note that in this study we used character space 
(See \suppref{Section~S4}{sec:char-space}) 
and hence the results are for any mode of the given symmetry in general, not just for some mode in particular.  It should also be noted that by using projection operators we are able to quantify the linear symmetry numerically, thus building upon the work of Ref.~\cite{Overvig:2020-35434:PRB} which gave a boolean value for the presence of the different types of linear symmetry.  Specifically, it is a new result that we are able to show the direction and gradient of change in linear symmetry as some parameter of the MS is changed.

\begin{figure}[htb]
    \centering
    \includegraphics[width=\columnwidth]{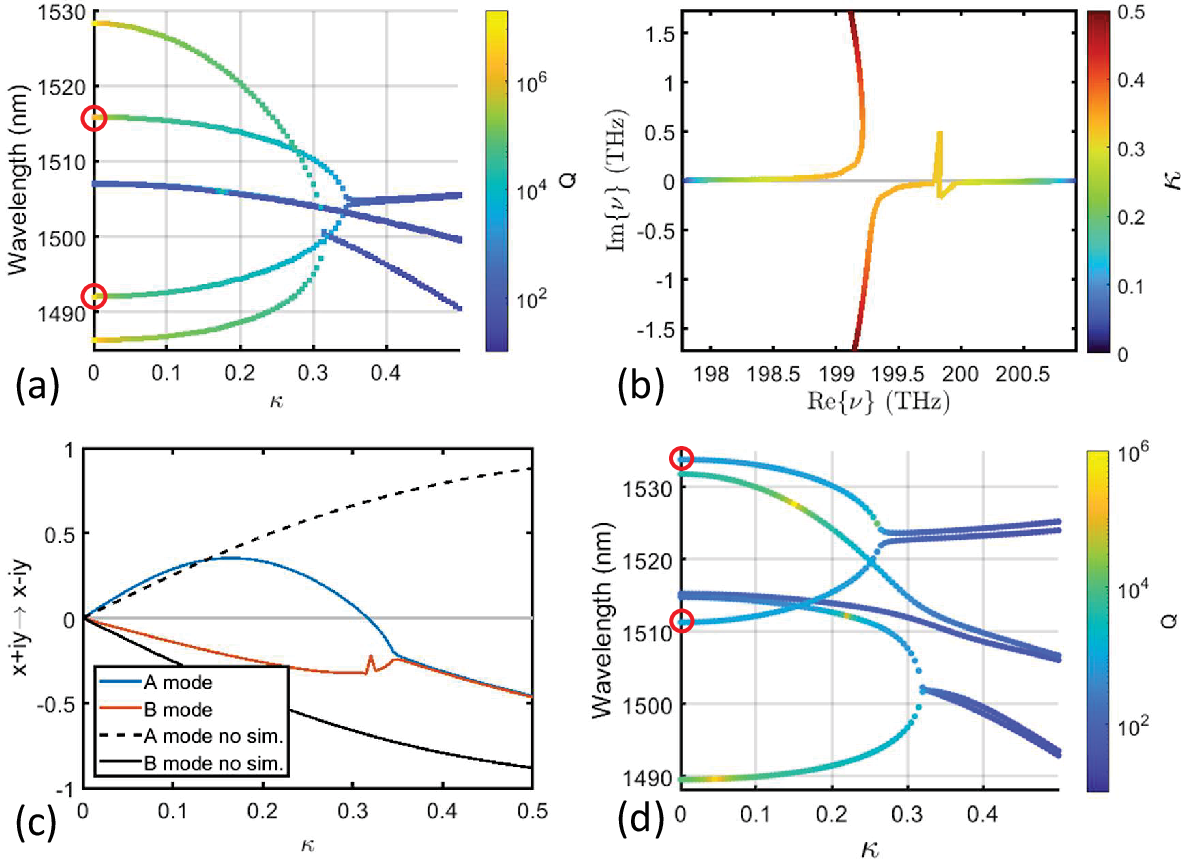}
	\caption[Fylfot metasurface with mode coupling]{
    (a) Eigenfrequency study for the case of the symmetric metasurface (no arm extension) with the TE modes indicated with red circles.  
    (b) Imaginary vs real part of the eigenfrequency for different values of \(\kappa\), the imaginary part of the refractive index. (c) Predictions for \(\hat{A}^{C4}\) and \(\hat{B}^{C4}\) mode linear symmetry, with comparison to the results in \Cref{fig:pt1}d. (d) Eigenfrequency study for the case of the asymmetric metasurface (arms extended by 50~nm) with the TE modes indicated with red circles.}
    \label{fig:pt3}
\end{figure}

To prove the presence of an EP, we simulated the symmetric design (no arm extension) for various values of \(\kappa\) with eigenfrequency studies (see \Cref{fig:pt3}a).  By restricting the study to the symmetric design, we avoid complicating mode couplings that we analyze later.  For this study, there is no substrate, the periodicity is 1400~nm, the height is 300~nm, the width of the arms is 150~nm, and the lengths of the arms are 400~nm and 150~nm for the longer and shorter sections.  The refractive index is \(3.3779\pm i\kappa\).  The two modes marked with red circles are the Transverse Electric (TE) \(\hat{A}^{C4}\) and \(\hat{B}^{C4}\) symmetric modes, with the other modes being Transverse Magnetic (TM).  As can be seen, at an EP, the two modes become completely coupled, as do the \(\hat{A}^{C4}\) and \(\hat{B}^{C4}\) TM modes, with the remaining line being the degenerate TM \(\hat{E}^{C4}\) modes.  In \Cref{fig:pt3}b, we can see the imaginary vs real parts of the eigenfrequencies of the TE modes, which also shows the modes reaching an EP at which one is gain and the other is matched loss. The short spike in the plot is where the \(\hat{B}^{C4}\) mode crosses the TM broken \(\mathcal{PT}\) phase modes.

We can now repeat the method of Ref.~\cite{Overvig:2020-35434:PRB} by using the simulations of the symmetric case, which will take into account some coupling between the modes.  The results of this study are shown in \Cref{fig:pt3}c, with the results of the previous study shown in black for comparison.  Because the previous study did not take into account any coupling between the modes, we would expect it to only be valid for low values of \(\kappa\), and this is borne out in the study.  Note also that after the EP, the complex linear symmetry is the same for both modes and again, they have no component in the real linear symmetry plane.

\subsection{Asymmetric Fylfot Metasurface}
We next simulated the same design as above but with the two arms extended by 50~nm, which remains within the first-order perturbation approximation required for the analysis of Ref.~\cite{Overvig:2020-35434:PRB}.  The dispersion of the eigenmodes for different values of \(\kappa\) is shown in \Cref{fig:pt3}d, where the two TE modes are marked with red circles.  The shorter wavelength mode is \(\hat{B}^{C4}\) symmetric, while the longer wavelength mode is \(\hat{A}^{C4}\) symmetric (See \suppref{Section~S9}{sec:mode-sym}). The very slight gap between the modes in the broken-\(\mathcal{PT}\) phase shows that the two modes are very slightly detuned from the exceptional point in this case. The other four modes are all four of the lowest-order TM modes. With coupling between four modes, the TM mode coupling would ordinarily be very difficult to analyze. Projection operators however make it is easy to show that the TM modes follow the same pattern as for the TE modes, but with an anti-crossing of one of the TM \(\hat{E}^{C4}\) modes with the TM \(\hat{A}^{C4}\) mode (See \suppref{Section~S9}{sec:mode-sym}).

\begin{figure}[thb]
    \centering
    \includegraphics[width=\columnwidth]{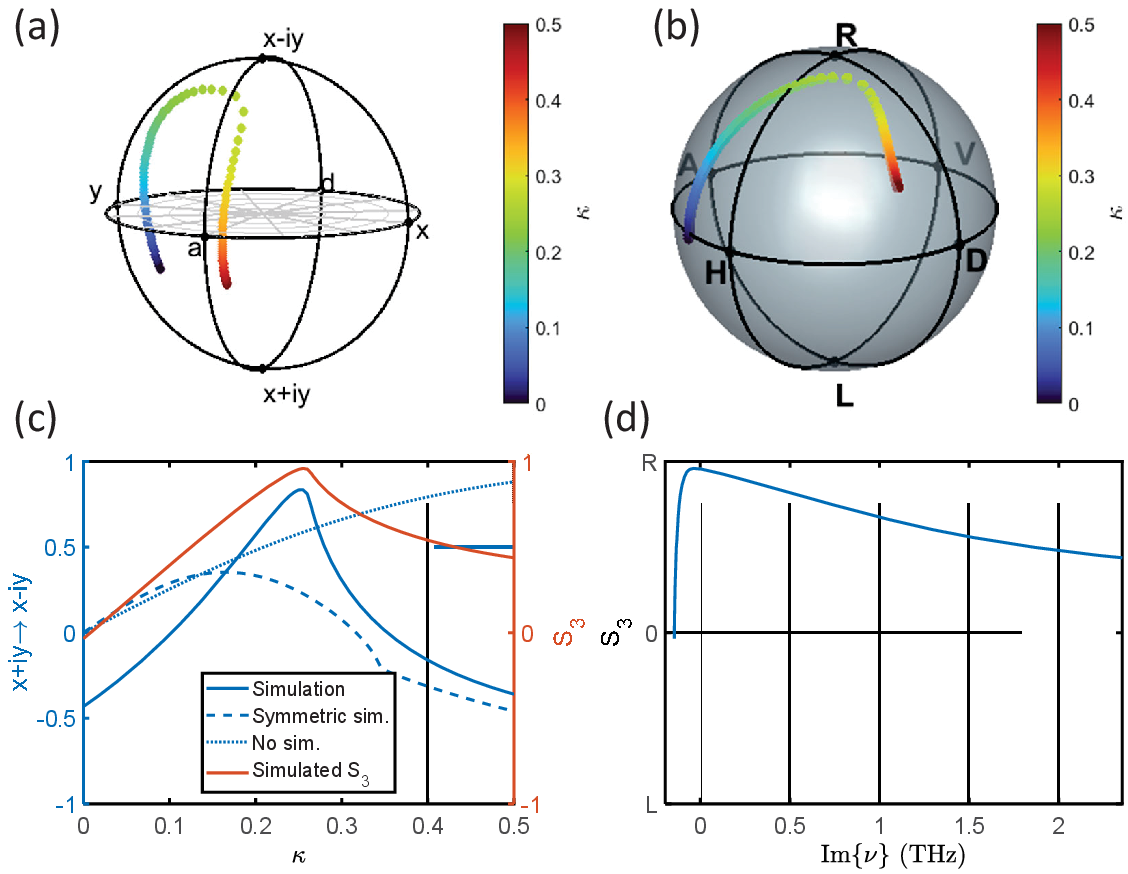}
	\caption[Fylfot metasurface: Full simulations]{
    The gain mode: (a) \(\hat{A}^{C4}\) mode symmetry with full simulations of metasurface with arms extended by 50~nm.  
    (b) Polarization in \(+z\) for the same mode.  
    (c) Mode symmetry found with this study compared to those of \Cref{fig:pt1,fig:pt3}. Stokes parameter \(S_3\) of the mode is also shown. 
    (d) Plot of \(S_3\) versus imaginary part of the eigenfrequency, showing complex polarization in the gain region.}
    \label{fig:pt4}
\end{figure}

In \Cref{fig:pt4}a, we can see the simulation results for the \(A\) mode, which takes into account the full mode couplings. The qualitative results of the previous studies are reproduced and, by comparing it to \Cref{fig:pt4}b, we can see that, as predicted, there is a strong correlation between the mode symmetry and the polarization.  This figure shows the eigenmode polarization at the \(+z\) port, while at the \(-z\) port, the polarization is the same but with the transformations \(S_3\rightarrow-S_3\) and \(S_2\rightarrow-S_2\).  In \Cref{fig:pt4}c, we can see all of the studies in the complex axis only, which shows the correspondences and differences as more mode coupling is taken into account. The red line shows the \(S_3\) axis of the mode's polarization. 

The important result here, though, is not whether or not the mode has complex polarization at some point but whether or not it is complex in the gain region. This is shown in \Cref{fig:pt4}d, which shows the polarization of the \(A\) mode as a function of the complex part of the eigenmode frequency. As can be seen, the mode polarization remains highly complex in the gain region. Further study is needed to determine whether the complexity can be increased by using optimization routines that use projectors to recast the analyses above in algorithmic form.

\subsection{Proposed Metasurface Laser}
Although it would be possible to use optimization routines to reduce the value of \(\kappa\) at which the transition to the broken \(\mathcal{PT}\) phase occurs, for an initial attempt it is important to make the MS as easy to manufacture as possible.  We have therefore opted for a simple design that, intuitively, one would expect to reduce the required value of \(\kappa\).  

\begin{figure}[tb]
    \centering
    \includegraphics[width=\columnwidth]{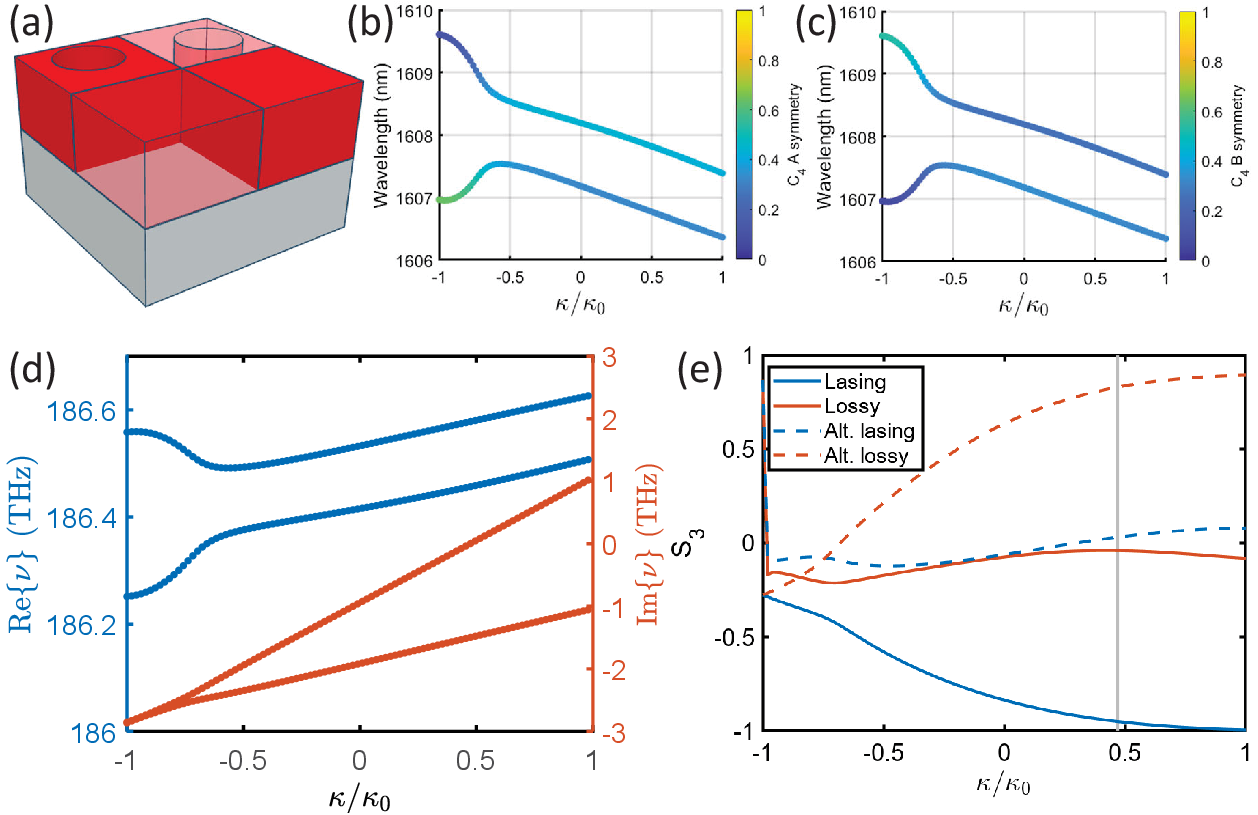}
	\caption[\(\mathcal{PT}\) metasurface]{
    (a) Design of the InGaAs on quartz MS: Periodicity is 1220~nm, height of InGaAs layer is 400~nm and radius of holes is 140~nm. The transparent regions are unpumped InGaAs and the opaque regions are pumped InGaAs so that the complex part of \(n=n_0-j\kappa\) is tuneable.  (b,c) \(C_4\ A\) (b) and \(C_4\ B\) (c) symmetry of of the two modes. (d) Real and imaginary parts of the eigenfrequency of the modes. (e) \(S_3\) axis of the polarization of the modes.}
    \label{fig:pt5}
\end{figure}

The MS consists of a uniform slab of InGaAs on a quartz substrate to which we add gain via patterned illumination.  To achieve the patterning, we angle four 532~nm laser beams at an angle of \(\approx26^\circ\) to the normal, since by assuming plane waves, we have in-plane wavevectors of
\begin{equation}
\frac{2\pi}{\lambda}\sin(\theta)=\frac{2\pi}{P}
\end{equation}
where the pitch is \(P=1220\)~nm.  Asymmetry is introduced by etching two holes in the InGaAs slab as shown in \Cref{fig:pt5}a.  The figure also shows that in the opaque region the InGaAs layer is pumped to obtain gain, and the transparent region is unpumped, lossy InGaAs. The periodicity is 1220~nm, with the gain regions being 610~nm square.  The height of the InGaAs layer is 400~nm, and the holes have a radius of 140~nm. Although in this case, due to the substrate, there is no z-axis up-down symmetry, the previous study (which did not have a substrate) showed that chiral emission is possible from a MS with up-down symmetry.

The gain region is from 1550~nm to 1650~nm, and we make the simplifying assumption that the first mode to have a positive complex part of the eigenfrequency will be the only mode to lase 
(See \suppref{Fig.~S5}{fig:lasing}). This assumption is a necessary, but not sufficient, condition for lasing.  
In \Cref{fig:pt5}b \&~c, we can see symmetry analyses of the lasing mode and its \(\mathcal{RT}\) symmetric partner in terms of the \(\hat{A}^{C4}\) and \(\hat{B}^{C4}\) symmetries.  Note that in these results we have used the standard of \(n=n_r-j\kappa_0\) for InGaAs, so that \(\flatfrac{\kappa}{\kappa_0}=-1\) corresponds to a uniform slab of InGaAs and \(\flatfrac{\kappa}{\kappa_0}=+1\) is matched gain and loss.  As in the previous analyses there are \(\hat{A}^{C4}\) and \(\hat{B}^{C4}\) modes, but they are no longer purely of these symmetries.  For \(\flatfrac{\kappa}{\kappa_0}>-0.5\), the two modes are not the same symmetry or at the same frequency so we can see that the \(\mathcal{PT}\) symmetry is detuned from the EP in this case.  In \Cref{fig:pt5}d, however, we can see that the main features of a \(\mathcal{PT}\) symmetric system are fulfilled by the two modes in the real and complex parts of the eigenfrequency.

Finally, in \Cref{fig:pt5}e, we can see the \(S_3\) axis of the polarization for the modes.  The blue solid line shows the lasing mode, and the orange solid line is the lossy \(\mathcal{RT}\) partner.  The figure shows that the lasing mode moves towards LHC polarization with increasing gain.  The gray line shows the point at which the mode becomes a gain mode and hence would begin lasing, where \(S_3<-0.95\).  At matched gain and loss we have \(S_3=-0.99\). We can also swap the gain and loss regions shown in \Cref{fig:pt5}a, which is not equivalent because of the asymmetry introduced by the holes.  From our coupled-mode theory we would expect the modes in this case to have the opposite polarization, which is exactly what we find in \Cref{fig:pt5}e.  The blue dashed line is the alternative lasing mode and the dashed orange line, shows the alternative lossy mode.  As can be seen, the lasing mode in the first configuration becomes the lossy mode in the alternative configuration and visa versa.  The alternative lasing mode begins lasing at the same value of \(\flatfrac{\kappa}{\kappa_0}=0.47\), and at matched gain and loss, we have \(S_3=0.08\).  We can thus see that the polarization can be actively switched from LCP to linear polarization by changing the illumination.  As noted above, these symmetry analyses could be included in an optimization routine to improve the result.

\section{Conclusion}
In conclusion, we have shown that projection operator based computational symmetry analysis provides an essential and efficient framework for symmetry control in photonic design and beyond, and can be integrated into optimization routines~\cite{Parry:2024-75406:PRB} or machine learning approaches. We have demonstrated the power of this method by applying it to planar metasurfaces engineered by symmetry principles to emit nearly pure circularly polarized light in the normal direction.
Armed with this tool, we introduced a manufacturable 
\(\mathcal{PT}\)-symmetric planar metasurface based on a patterned InGaAs slab on quartz, 
where the mode that first reaches a positive imaginary part of the eigenfrequency is a lasing candidate 
whose emission can be actively tuned between LCP and linear polarization by adjusting the illumination.


\emph{Acknowledgments} -  We acknowledge the support from the Australian Research Council via the Centres of Excellence (CE200100010) and Future Fellowships (FT230100058) programs.


\nocite{fraleigh1987linear,Parry:2024-75406:PRB,projop,McWeeny1963,Reinke:2011-66603:PRE,Willock2009,Schattschneider1978,Carter2009,clamv,Sakoda:2004:OpticalProperties}

\bibliography{bibliography}

\end{document}